\documentclass[sigconf]{acmart}
\usepackage{multirow}
\usepackage{graphicx}
\usepackage{xspace}
\usepackage{amsthm}
\usepackage{amsmath}
\usepackage{enumitem}
\usepackage{booktabs}
\usepackage{tcolorbox}
\usepackage{float}
\usepackage{caption}
\usepackage{subcaption}
\usepackage{balance}

\usepackage[toc,page]{appendix}

\AtBeginDocument{%
  \providecommand\BibTeX{{%
    \normalfont B\kern-0.5em{\scshape i\kern-0.25em b}\kern-0.8em\TeX}}}

\copyrightyear{2025}
\acmYear{2025}
\setcopyright{acmcopyright}
\acmDOI{XXXXXXX.XXXXXXX}

\acmConference[SIGIR'25]{Make sure to enter the correct
  conference title from your rights confirmation emai}{July 13--18, 2025}{Padova, Italy}
\acmPrice{15.00}
\acmISBN{978-1-4503-XXXX-X/18/06}

\begin{document}
\title[AGP]{Automating Personalization: Prompt Optimization for Recommendation Reranking}

\author{Chen Wang}
\authornote{Equal Contributions}
\orcid{0000-0001-5264-3305}
\affiliation{%
  \institution{University of Illinois Chicago}
  \streetaddress{...}
  \city{Chicago}
  \state{Illinois}
  \country{USA}
  \postcode{60607}
}
\email{cwang266@uic.edu}

\author{Mingdai Yang}
\orcid{0000-0002-2868-8965}
\authornotemark[1]
\affiliation{%
  \institution{The University of Chicago}
  \city{Chicago}
  \state{Illinois}
  \country{USA}
  \postcode{60607}
}
\email{frankyang@uchicago.edu}

\author{Zhiwei~Liu}
\authornote{Corresponding author}
\orcid{0000-0003-1525-1067}
\affiliation{%
  \institution{Salesforce AI Research}
  \city{Palo Alto}
  \state{California}
  \country{USA}
}
\email{zhiweiliu@salesforce.com}

\author{Pan~Li}
\affiliation{%
  \institution{Georgia Tech}
  \city{Atlanta}
  \state{Georgia}
  \country{USA}
}
\email{pan.li@scheller.gatech.edu}

\author{Linsey~Pang}
\affiliation{%
  \institution{Salesforce AI Research}
  \city{Palo Alto}
  \state{California}
  \country{USA}
}
\email{panglinsey@gmail.com}

\author{Qingsong~Wen}
\affiliation{%
  \institution{Squirrel Ai Learning}
  \state{California}
  \country{USA}
}
\email{qingsongedu@gmail.com}

\author{Philip Yu}
\affiliation{%
  \institution{The University of Chicago}
  \city{Chicago}
  \state{Illinois}
  \country{USA}
  \postcode{60607}
}
\email{psyu@uic.edu}

\renewcommand{\shortauthors}{Chen Wang et al.}

\begin{abstract}
Modern recommender systems increasingly leverage large language models (LLMs) for reranking to improve personalization. However, existing approaches face two key limitations: (1) heavy reliance on manually crafted prompts that are difficult to scale, and (2) inadequate handling of unstructured item metadata that complicates preference inference. We present \textbf{AGP} (Auto-Guided Prompt Refinement), a novel framework that automatically optimizes user profile generation prompts for personalized reranking. AGP introduces two key innovations: (1) position-aware feedback mechanisms for precise ranking correction, and (2) batched training with aggregated feedback to enhance generalization. Extensive experiments across three diverse datasets (\textit{Amazon Movies \& TV}, \textit{Yelp}, and \textit{Goodreads}) demonstrate AGP's effectiveness, achieving improvements of 5.61--20.68\% in NDCG@10 over baseline models with just 100 training users. Our results show AGP's particular strength in enhancing graph-based recommenders (9.36--20.68\% gains for LightGCN) while maintaining strong performance with sequential models.
\end{abstract}

\ccsdesc[500]{Information systems~Recommender systems}


\keywords{Prompt Optimization, Recommender Systems, Large Language Models (LLMs), Collaborative Filtering, Reranking
}


\maketitle

\section{Introduction}
\emph{Reranking}, which refines initial recommendations to better align with user preferences, plays a pivotal role in improving recommendation quality~\cite{pei2019personalized, li2019personalised, lin2024discrete}. Recent advancements in large language models (LLMs) have shown promise for reranking tasks by capturing complex user interests via contextual understanding~\cite{zhang2025enhancing, luo2024recranker, carraro2024enhancing,zhang2024ai}. 

However, their effectiveness relies heavily on manually crafting prompts -- a labor-intensive process that demands significant domain expertise and limits scalability. Moreover, manually designed prompts struggle to address the complexity and diversity of user preferences. For instance, crafting prompts to capture nuanced user interests from user-item interactions, such as item titles or descriptions, often requires iterative trial-and-error. This process is not only time-consuming but also prone to suboptimal results due to its reliance on intuition rather than systematic optimization. Moreover, static prompts fail to adapt to dynamic datasets and evolving user behaviors, limiting their ability to deliver personalized recommendations at scale.

Existing research on prompt optimization largely focuses on tasks like question answering~\cite{sabbatella2024prompt}, mathematical reasoning~\cite{shao2024deepseekmath}, and news recommendation~\cite{recprompt}. 
\textit{RecPrompt}~\cite{recprompt} introduces a self-tuning prompting framework incorporating TopicScore to enhance explainability in news recommendations. However, this approach relies on structured content and topical consistency, making it less applicable to heterogeneous item recommendation scenarios, where item metadata can be inconsistent, unstructured, and user-generated. Current optimization methods~\cite{recprompt, gao2024llm} usually rely on aggregated ranking metrics like AUC or NDCG, which are useful for performance evaluation but insufficient for direct optimization guidance. Input reranking approaches~\cite{dehghankar2024rank} reorder items based on relevance and exposure but do not offer structured feedback to refine user preference modeling. A more interpretable and systematic strategy is needed to close the gap between LLM-based reranking and explicit user feedback.

To address these challenges, in this paper we propose \textbf{AGP: Auto-Guided Prompt Refinement for Personalized Reranking}, a novel framework that optimizes \textit{user profile generation prompts} rather than directly modifying reranking prompts. By refining user profiles, AGP enables LLMs to better capture personalized interests, leading to more effective and explainable reranking. AGP improves optimization through a \textbf{batched training with batched feedback} mechanism. Instead of refining prompts on a per-user basis, AGP evaluates multiple users simultaneously, preventing overfitting to individual cases and enhancing generalization. During each iteration, AGP systematically analyzes why a generated user profile fails to prioritize ground-truth items, ensuring that refinements are both meaningful and robust. To further enhance optimization, AGP introduces \textbf{position-based feedback}, which explicitly signals ranking misalignment. Unlike traditional ranking metrics such as NDCG, which serve as indirect optimization objectives, position-based feedback provides actionable instructions by identifying discrepancies between predicted and ideal item rankings. For example, if an item is ranked 3rd but should be 1st, AGP generates structured feedback to adjust the user profile generation prompt accordingly. This feedback is then aggregated across batches, allowing learned refinements to generalize across different users. AGP’s optimization process follows an \textbf{iterative batch-based strategy}, where feedback is collected over multiple iterations. Each batch generates actionable instructions that refine the user profile prompt in a gradient-like manner, ensuring continuous improvement without overfitting. By integrating structured user profile optimization with position-aware feedback, AGP enables a more interpretable and scalable approach to personalized reranking.

\paragraph{Our Contributions.}
\begin{itemize}[leftmargin=*]
    \item \textbf{User Profile Generation Prompt.} We propose refining a \emph{user profile generation prompt}---as opposed to a single-step rerank prompt---to more effectively capture user preferences from noisy textual data. This design enables more nuanced and robust personalization.
    \item \textbf{Position-Based Feedback.} We introduce a \emph{position-based feedback} mechanism that pinpoints item-level ranking misalignments, providing interpretable signals for iterative refinement. This approach proves more actionable than relying solely on aggregated metrics like NDCG.
    \item \textbf{Batched Training and Summarized Feedback.} We devise a \emph{batched} optimization framework that aggregates feedback across multiple users, mitigating overfitting to individual quirks and ensuring updates remain broadly applicable.
\end{itemize}
\section{Methodology}
\label{sec:methodology}
In this section, we formalize the reranking task, introduce our Auto Prompt Optimization Framework, and detail each of its components. We first define the problem setup and notation, then elaborate on the user profile generation process, the evaluation and feedback mechanism, and finally the iterative prompt optimization strategy.

\begin{figure}
    \centering
    \includegraphics[width=\linewidth]{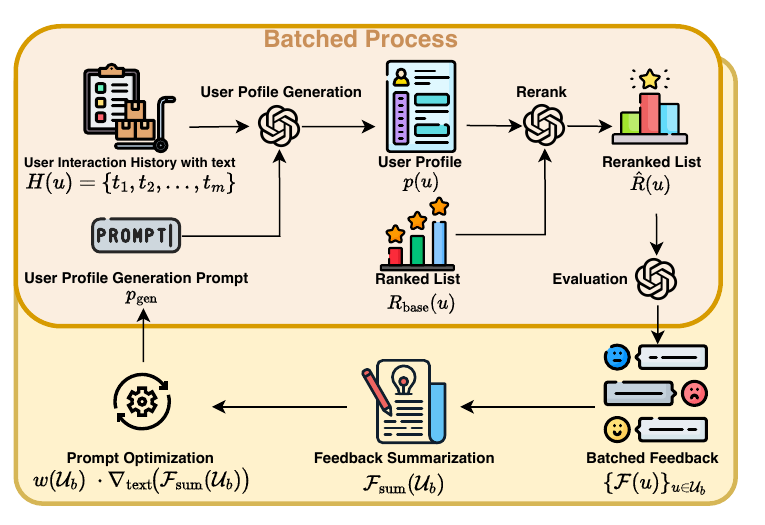}
    \caption{The pipeline of the AutoGuidePrompt (AGP) framework, illustrating the process from user interaction history to optimized reranked lists.}
    \label{fig:intro}
\end{figure}

\subsection{Problem Setup and Notation}
\label{sec:problem_setup}

Given a set of users $\mathcal{U}$, each user $u \in \mathcal{U}$ has an interaction history $H(u) = \{t_1, t_2, \ldots, t_m\}$, where each $t_j$ is the \textit{textual title} of an interacted item. Unlike ID-based approaches, this representation leverages semantic knowledge, enabling LLMs to infer preferences.

A recommender system provides a \textit{baseline ranking} $R_{\text{base}}(u) = \{i_1, i_2, \ldots, i_k\}$, where items are ranked by predicted relevance. The goal is to refine $R_{\text{base}}(u)$ into an optimized ranking $R_{\text{LLM}}(u)$ that better aligns with user preferences, and the reranking function is:
\begin{equation}
f: (\mathcal{U}, R_{\text{base}}) \rightarrow R_{\text{LLM}}.
\end{equation}

\paragraph{Challenges.} Applying LLMs to item recommendations presents two key challenges:
\begin{itemize}[leftmargin=*]
    \item \textbf{Disparate and Noisy Item Information:} Unlike structured domains like news recommendation, item representations in recommender systems vary widely, with metadata that can be \textbf{incomplete, inconsistent, or redundant} (e.g., short titles, varying descriptions, or noisy user-generated content). This variability makes it difficult for LLMs to infer structured user preferences.
    \item \textbf{Lack of Direct Optimization Signals:} Existing methods (e.g., RecPrompt) optimize prompts using aggregated ranking metrics (AUC, NDCG), which measure performance but do not directly provide optimization guidance. A more interpretable strategy is needed to refine prompts using explicit ranking signals.
\end{itemize}

To address these, we propose \textbf{AGP}, which optimizes \textit{user profile generation prompts} instead of reranking prompts, incorporates structured feedback, and iteratively refines prompts to improve personalization and ranking quality.

\subsection{User Profile Generation with a Learned Prompt}
\label{sec:user_profile_generation}

AGP optimizes a shared \emph{profile-generation prompt} $p_{\mathrm{gen}}$ to construct personalized user profiles. Unlike manually crafted prompts, $p_{\mathrm{gen}}$ is iteratively refined through batch training to capture generalizable patterns across users.

For each user $u$, AGP generates a profile based on two inputs: (1) the user’s text-based interaction history $H(u)$ and (2) the current version of $p_{\mathrm{gen}}$. The LLM synthesizes these inputs to generate a structured profile:
\begin{equation}
p(u) = \text{LLM}\bigl(H(u),\, p_{\mathrm{gen}}\bigr).
\end{equation}
Since $H(u)$ consists of item titles, the LLM extracts thematic preferences (e.g., \textit{science fiction, deep learning}) while $p_{\mathrm{gen}}$ provides structure and emphasis.

The generated profile $p(u)$ is then used to \emph{rerank} the baseline recommendation list $R_{\text{base}}(u)$, where the LLM refines rankings as:
\begin{equation}
\hat{R}(u) = \text{LLM}\bigl(p(u),\, R_{\text{base}}(u)\bigr).
\end{equation}
This two-step process enhances LLM reasoning by summarizing user interests before reranking, allowing informed and context-aware ranking decisions. Since AGP refines profile generation rather than modifying raw rankings, it preserves user-specific insights while generalizing effectively across different users.

\subsection{Position-Based Evaluation and Feedback}
\label{sec:evaluation_feedback}

To refine reranking, we introduce \emph{position-based feedback}, a more interpretable alternative to aggregated metrics like NDCG. Given a user $u$, we define a set of ground-truth relevant items $G(u) \subseteq \hat{R}(u)$ in the reranked list $\hat{R}(u)$. For each item $i \in G(u)$, we record its actual position $r_{\hat{R}}(i,u)$ and compare it to its target position $r_{\text{target}}(i,u)$. If an item should ideally be ranked in the top-3 but appears at position 5, the LLM receives a correction signal.

This process generates feedback signals:
\begin{equation}
\mathcal{F}(u) = \bigl\{ \bigl(r_{\hat{R}}(i,u),\, r_{\text{target}}(i,u)\bigr) \;\big|\; i \in G(u)\bigr\}.
\end{equation}
These signals specify ranking deviations, providing direct and interpretable instructions for refinement. Unlike aggregated ranking metrics, which provide an overall evaluation score, position-based feedback delivers explicit corrections for each item, making adjustments more targeted.

Position-based feedback offers two key advantages. First, it improves interpretability by providing explicit positional corrections rather than relying on abstract scores. Second, it enables fine-grained ranking adjustments, ensuring that high-relevance items receive stronger refinements while less critical items undergo minor tweaks. This process naturally integrates with AGP’s optimization strategy by offering direct ranking signals that guide systematic improvements.

\subsection{Batch Formation and Optimization}
\label{sec:prompt_optimization}

AGP optimizes the profile-generation prompt $p_{\mathrm{gen}}$ using batch training. In each iteration, a batch $\mathcal{U}_b$ of users is processed, where each user $u \in \mathcal{U}_b$ generates a profile $p(u) = \text{LLM}\bigl(H(u),\, p_{\mathrm{gen}}\bigr)$, which is then used to compute the reranked list $\hat{R}(u)$. Feedback $\mathcal{F}(u)$, derived from position-based evaluation (Sec.~\ref{sec:evaluation_feedback}), guides prompt refinement.

Batch-based training has been previously explored for improving ranking robustness~\cite{dehghankar2024rank}, where partitioned input evaluation helps mitigate bias. Inspired by this, AGP extends batch training to structured prompt optimization by aggregating feedback across users, ensuring refinements are driven by explicit ranking misalignments rather than indirect scoring metrics. Given individual position-based signals $\mathcal{F}(u)$, AGP generates a summarized set of batch-level improvements:
\begin{equation}
\mathcal{F}_{\mathrm{sum}}(\mathcal{U}_b) = \sum_{u \in \mathcal{U}_b} w(u) \mathcal{F}(u),
\end{equation}
where $w(u) = \frac{1}{\text{avgPos}(u)}$ assigns a higher weight to users whose ground-truth items rank lower on average.

The prompt is iteratively updated as:
\begin{equation}
p_{\mathrm{gen}} \;\leftarrow\; p_{\mathrm{gen}} - w(\mathcal{U}_b) \cdot \nabla_{\mathrm{text}}\!\bigl(\mathcal{F}_{\mathrm{sum}}(\mathcal{U}_b)\bigr),
\end{equation}
where $w(\mathcal{U}_b) = \frac{1}{\text{avgPos}(\mathcal{U}_b)}$ ensures stronger updates when ranking errors are larger.

This iterative process refines $p_{\mathrm{gen}}$ by adjusting user profile descriptions based on ranking misalignment. After each update, new user profiles $p(u)$ and reranked lists $\hat{R}(u)$ are generated, and feedback is recalculated until convergence. By summarizing batch-wide trends, AGP prevents overfitting to specific user preferences and ensures adaptability across diverse user populations.

\section{Experiments}
\label{sec:experiments}
\subsection{Experimental Settings}
\label{sec:experimental_setup}
We use three public datasets: \textit{Amazon Movies \& TV} (Movies), \textit{Yelp}, and \textit{Goodreads}. The Amazon dataset has 95,593 users, 43,117 items, and 750,081 interactions; Yelp has 65,348 users, 33,626 items, and 1,041,540 interactions; and Goodreads has 13,100 users, 25,434 items, and 856,280 interactions. Using a leave-one-out (LOO) strategy, we designate the most recent user interaction as the test item and the second-most recent as validation. To establish a baseline, we train two recommender models: \textbf{(1) LightGCN}~\cite{he2020lightgcn}, a graph-based collaborative filtering method, and \textbf{(2) SASRec}~\cite{kang2018self}, a sequential transformer-based recommender. Each model generates top-10 predictions per user, from which we randomly select 300 users for evaluation, ensuring each ground-truth item is included. On this subset, we apply our AGP framework to \emph{rerank} the top-10 list. AGP is trained on 100 randomly selected user interaction histories with a maximum of 10 epochs. We explore user history sequence lengths of 5, 10, and 20 as a hyperparameter, along with batch size values of 5, 10, and 20. Evaluation is conducted using \textbf{NDCG@10}, which measures ranking quality (higher is better). We also compare against two non-iterative baselines: \emph{LLM-Dir} (single-pass prompt for reranking) and \emph{LLM-CoT} (single-pass chain-of-thought prompt~\cite{kojima2022large} for reranking). Additionally, we include four LLMs for performance comparison: GPT-4o (4o), GPT-4o-Mini (4o-Mini), GPT-o3-Mini (o3-Mini), and DeepSeek-V3 (DeepSeek)~\cite{deepseekai2024deepseekv3technicalreport}.

\subsection{Results and Discussion}

The results are shown in Table~\ref{tab:rerank_results}, highlighting key findings. \textbf{AGP effectiveness in item recommendation:} AGP proves effective for reranking tasks, demonstrating that LLMs can self-optimize prompts within our framework. This adaptability reduces the need for manual prompt tuning and enhances ranking quality, particularly in \textbf{personalized recommendation scenarios}. \textbf{LLM reranker performance on SASRec vs. LightGCN:} LLM rerankers show greater improvements on SASRec rankings than on LightGCN. This is likely due to SASRec and LLMs both leveraging \textbf{time-series modeling}, making them more effective in capturing sequential user behaviors. LightGCN, a \textbf{graph-based collaborative filtering model}, focuses on global user-item interactions, which limits LLMs' ability to enhance its ranking list. \textbf{CoT effectiveness:} The best performance on the Yelp dataset is achieved by o3-Mini, indicating that in scenarios with \textbf{insufficient textual context and noisy data}, multi-step reasoning is beneficial. This suggests that structured thinking models like o3-Mini can extract better signals in complex and sparse text environments, improving reranking effectiveness. \textbf{Yelp dataset challenges:} The Yelp dataset sees minor improvements due to the \textbf{lack of rich textual features and presence of noisy text}. Many business descriptions are sparse or inconsistent, making it difficult for LLMs to infer meaningful item relationships. Despite these challenges, AGP remains effective by dynamically refining prompts, allowing LLMs to better interpret available textual information and improve reranking performance.

\begin{table}[h]
\centering
\small 
\setlength{\tabcolsep}{4pt} 
\renewcommand{\arraystretch}{0.9} 
\caption{Reranking performance (N@10) across datasets. ``T100'' denotes AGP trained on 100 users. \textbf{Bold} values indicate the best performance within the same LLM model.}
\label{tab:rerank_results}
\begin{tabular}{llccc}
\toprule
\multirow{2}{*}{\textbf{Model}} & \multirow{2}{*}{\textbf{Method}}  
& \multicolumn{3}{c}{\textbf{N@10}}  \\
\cmidrule(lr){3-5}
& & \textbf{AMZ} & \textbf{YELP} & \textbf{GR} \\
\midrule
\multirow{14}{*}{\centering \emph{LightGCN}}  
& Base & 0.513 & 0.501 & 0.474 \\
\cmidrule(lr){2-5}
& + LLM-Dir (4o) & 0.547 & 0.491 & 0.555 \\
& + LLM-CoT (4o) & 0.551 & 0.491 & 0.560 \\
& + AGP-T100 (4o) & \textbf{0.553} & \textbf{0.502} & \textbf{0.572} \\
\cmidrule(lr){2-5}
& + LLM-Dir (4o-Mini) & 0.553 & 0.484 & 0.548 \\
& + LLM-CoT (4o-Mini) & 0.553 & 0.481 & 0.547 \\
& + AGP-T100 (4o-Mini) & \textbf{0.561} & \textbf{0.493} & \textbf{0.553} \\
\cmidrule(lr){2-5}
& + LLM-Dir (o3-Mini) & 0.542 & 0.538 & 0.540 \\
& + LLM-CoT (o3-Mini) & 0.543 & 0.531 & 0.542 \\
& + AGP-T100 (o3-Mini) & \textbf{0.558} & \textbf{0.541} & \textbf{0.548} \\
\cmidrule(lr){2-5}
& + LLM-Dir (DeepSeek) & 0.546 & 0.496 & 0.533 \\
& + LLM-CoT (DeepSeek) & 0.547 & 0.498 & 0.544 \\
& + AGP-T100 (DeepSeek) & \textbf{0.551} & \textbf{0.504} & \textbf{0.564}  \\
\midrule
\multirow{14}{*}{\centering \emph{SASRec}}  
& Base & 0.659 & 0.528 & 0.599 \\
\cmidrule(lr){2-5}
& + LLM-Dir (4o) & 0.671 & 0.510 & 0.624 \\
& + LLM-CoT (4o) & 0.664 & 0.521 & 0.610 \\
& + AGP-T100 (4o) & \textbf{0.696} & \textbf{0.530} & \textbf{0.636} \\
\cmidrule(lr){2-5}
& + LLM-Dir (4o-Mini) & 0.654 & 0.502 & \textbf{0.631} \\
& + LLM-CoT (4o-Mini) & 0.656 & 0.507 & 0.615\\
& + AGP-T100 (4o-Mini) & \textbf{0.658} & \textbf{0.517} & 0.627\\
\cmidrule(lr){2-5}
& + LLM-Dir (o3-Mini) & 0.658 & 0.527 & 0.587\\
& + LLM-CoT (o3-Mini) & 0.663 & 0.528 & 0.585\\
& + AGP-T100 (o3-Mini) & \textbf{0.683} & \textbf{0.541} & \textbf{0.595} \\
\cmidrule(lr){2-5}
& + LLM-Dir (DeepSeek) & 0.648 & 0.512 & 0.622 \\
& + LLM-CoT (DeepSeek) & 0.655 & 0.519 & 0.610 \\
& + AGP-T100 (DeepSeek) & \textbf{0.687} & \textbf{0.523} & \textbf{0.636} \\
\bottomrule
\end{tabular}
\end{table}

\subsection{Ablation Study}

To evaluate the impact of design choices in our framework, we conduct an ablation study using GPT-4o on the AMZ dataset, focusing on three key aspects: the effect of summarization in reducing overfitting, the influence of batch size and sequence length on reranking performance, and the effectiveness of position-based feedback.

\begin{figure}[t]
    \centering
    \begin{subfigure}{0.48\columnwidth}
        \centering
        \includegraphics[width=\linewidth]{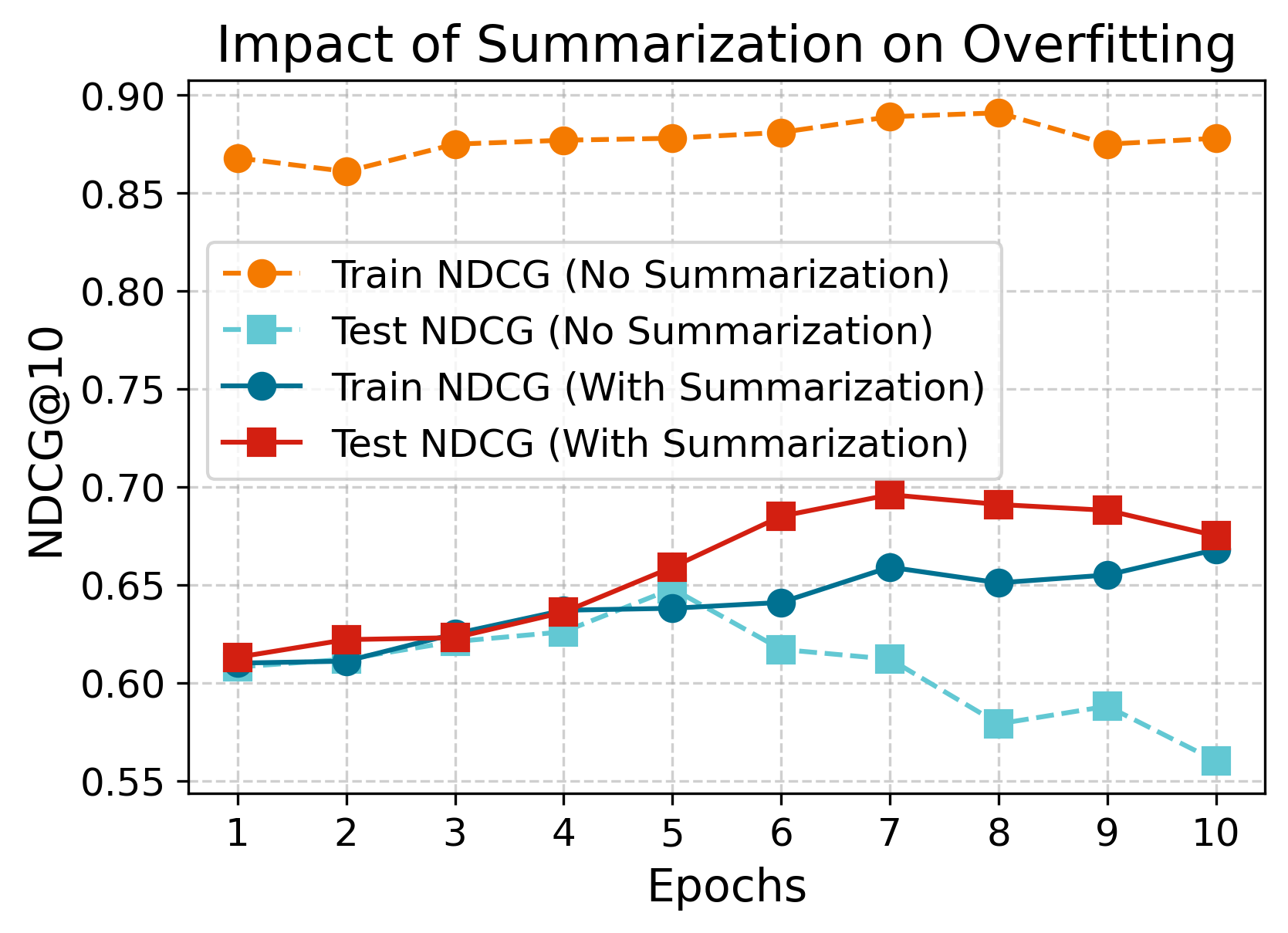}
    \end{subfigure}
    \hfill
    \begin{subfigure}{0.48\columnwidth}
        \centering
        \includegraphics[width=\linewidth]{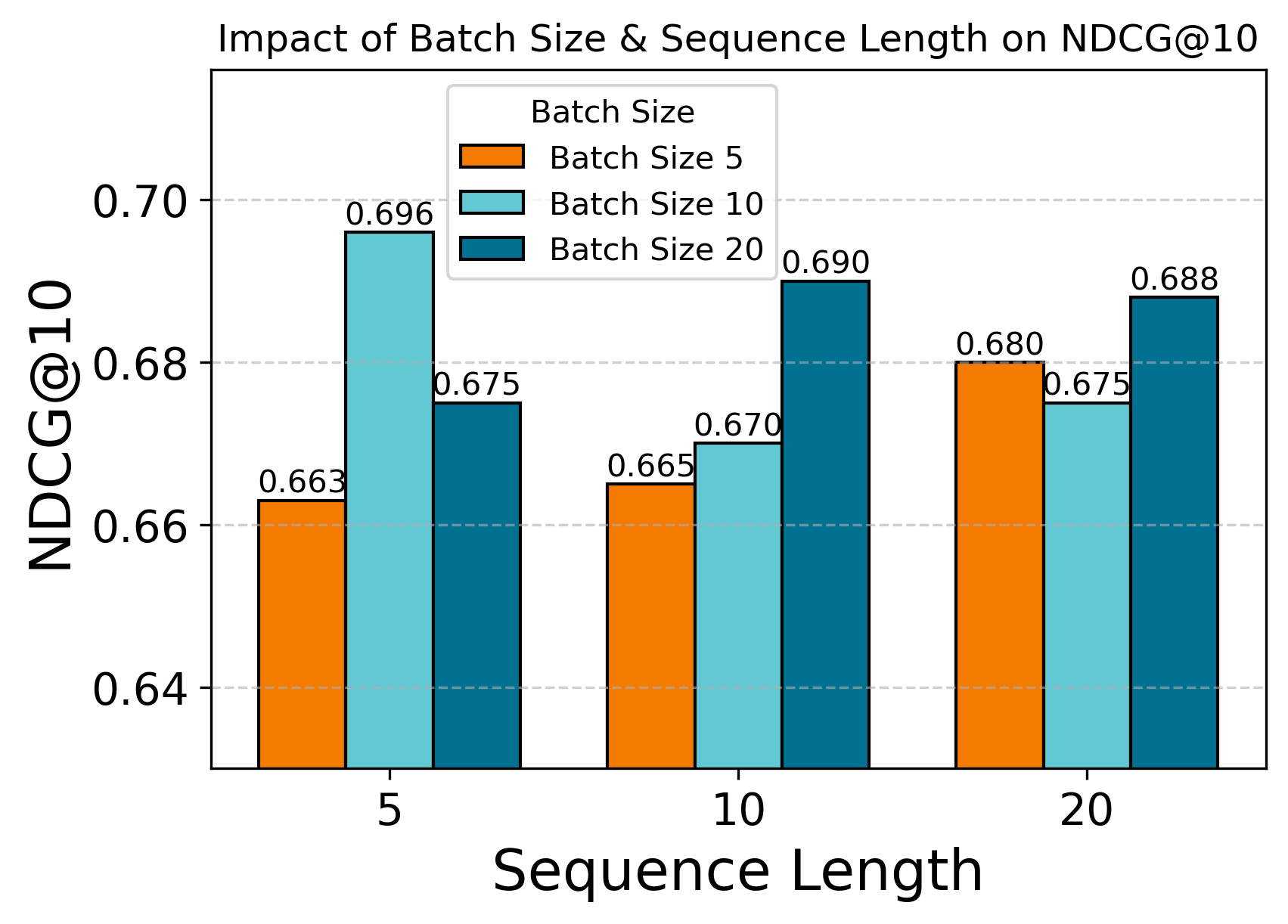}
    \end{subfigure}
    \caption{Ablation study on summarization (left) and batch size/sequence length impact (right) using GPT-4o on the AMZ dataset. Summarization reduces overfitting, while batch size 10 and sequence length 5 yield optimal ranking performance.}
    \label{fig:ablation_study}
\end{figure}

\begin{figure}[t]
    \centering
    \begin{subfigure}{0.32\columnwidth}
        \centering
        \includegraphics[width=\linewidth]{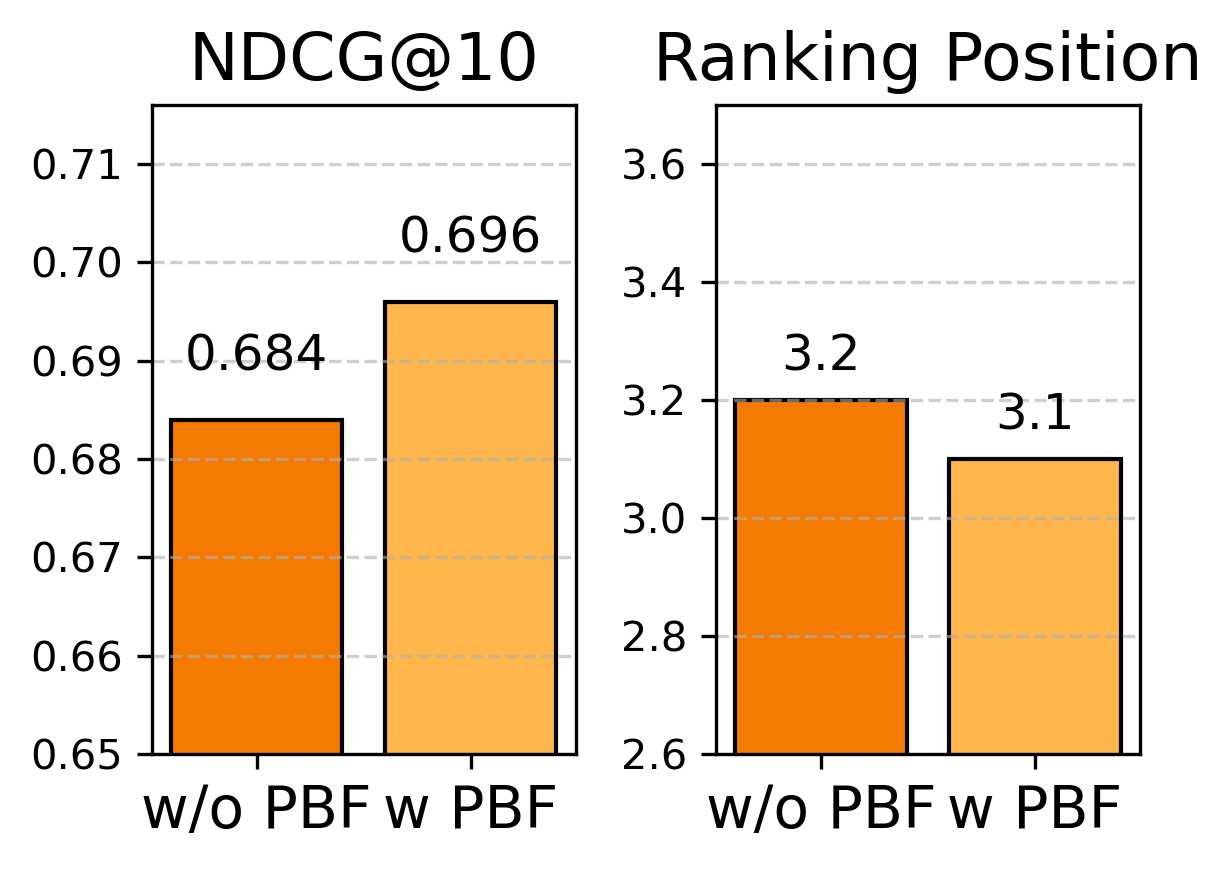}
        \caption{Amazon Dataset}
        \label{fig:ablation_amz}
    \end{subfigure}
    \hfill
    \begin{subfigure}{0.32\columnwidth}
        \centering
        \includegraphics[width=\linewidth]{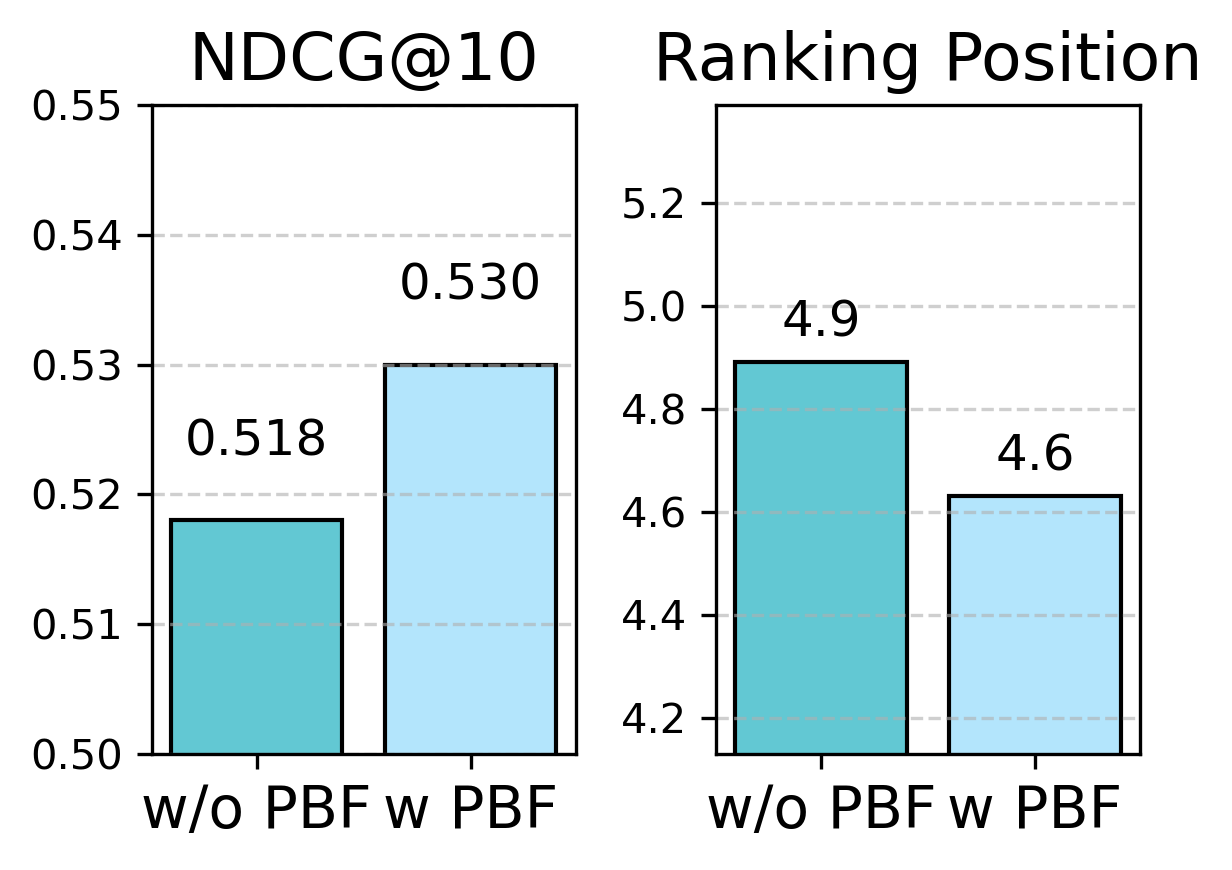}
        \caption{Yelp Dataset}
        \label{fig:ablation_yelp}
    \end{subfigure}
    \hfill
    \begin{subfigure}{0.32\columnwidth}
        \centering
        \includegraphics[width=\linewidth]{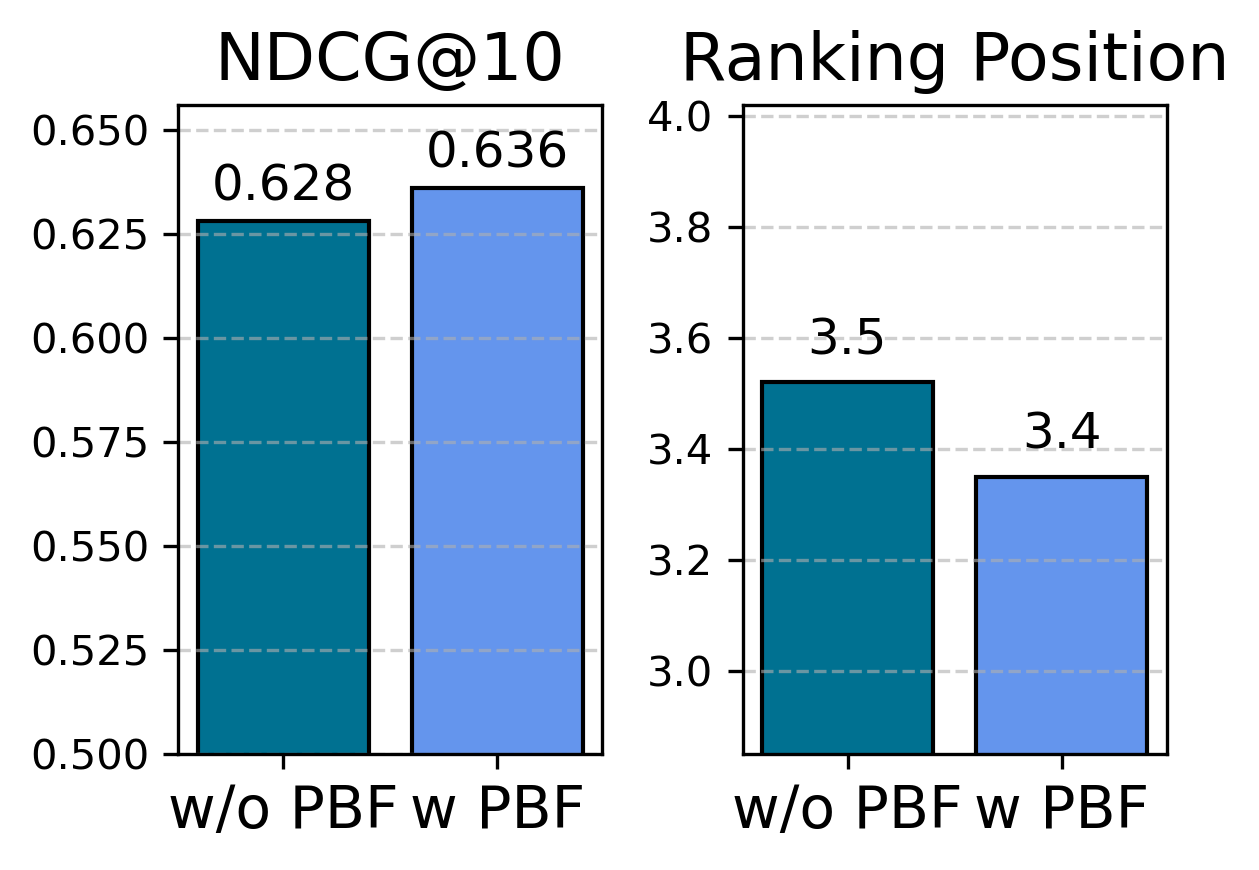}
        \caption{Goodreads Dataset}
        \label{fig:ablation_gr}
    \end{subfigure}
    \caption{Ablation study on Position-Based Feedback.}
    \label{fig:ablation_study1}
\end{figure}

\textbf{Summarization Impact on Overfitting:} We analyze the effect of summarization by comparing training and test performance with and without summarization. As shown in Figure~(\ref{fig:ablation_study}, left), summarization prevents excessive fitting to the training data while improving test performance, enhancing generalization by filtering redundant information and refining prompt optimization. \textbf{Effectiveness of Batch Size and Sequence Length:} We examine how batch size and sequence length impact reranking effectiveness. Figure~(\ref{fig:ablation_study}, middle) shows that the best performance is achieved with a batch size of 10 and a sequence length of 5, suggesting an optimal balance between convergence and generalization. Larger batch sizes stabilize training, while shorter sequences reduce noise from long interaction histories, highlighting the importance of careful hyperparameter selection for improving LLM reranking. \textbf{Effectiveness of Position-Based Feedback (PBF):} We further investigate the impact of PBF on reranking quality across datasets. Figure~\ref{fig:ablation_study1} demonstrates that incorporating PBF improves both \textit{NDCG@10} and average ranking position across all datasets. The gains are more significant in datasets with high variability in ranking scores, indicating that PBF helps LLMs better capture relative item importance. This result highlights its potential in enhancing ranking stability and improving personalization in LLM-based reranking systems.

\subsection{Training Efficiency}
We evaluate the efficiency of AGP training by analyzing the API call calculations and performance scaling. The total API calls per training stage follow the formula:
\begin{equation}
    \text{API Calls} = (\text{batch\_size} \times 3 + 2) \times \frac{100}{\text{batch\_size}}
\end{equation}
where 3 corresponds to generating a user profile, reranking, and computing the loss per user, while 2 accounts for summarization and prompt optimization per batch. To further assess AGP’s efficiency, we trained it on the AMZ dataset with GPT-4o using 700 users instead of 100. The NDCG@10 increased marginally from 0.696 to 0.705 (a 1.29\% increase), demonstrating that AGP maintains competitive performance with significantly fewer training samples. This result underscores AGP’s effectiveness and efficiency, reducing computational costs while preserving high reranking quality.

\section{Conclusion}
We propose \textbf{AGP}, a framework optimizing \textit{user profile generation prompts} for better LLM-based reranking. reduce one word: AGP employs \textbf{batched feedback} and \textbf{position-based feedback} for improved generalization and ranking accuracy. Experiments confirm its effectiveness, with future work exploring reinforcement learning and broader use.

\bibliographystyle{ACM-Reference-Format}
\balance
\bibliography{sample-base}

\end{document}